\documentclass[twocolumn,showpacs,preprintnumbers,amsmath,amssymb]{revtex4}

\usepackage{graphicx}
\usepackage{dcolumn}
\usepackage{bm}
\usepackage{psfig}
\usepackage{longtable}

\begin{document}


\title{Three-body Thomas-Ehrman shifts of analog states of $^{17}$Ne 
and $^{17}$N}

\author{E. Garrido} 
\email{imteg57@pinar2.csic.es}
\affiliation{ Instituto de Estructura de la Materia, CSIC, 
Serrano 123, E-28006 Madrid, Spain }

\author{D.V. Fedorov}
\author{A.S. Jensen}
\affiliation{ Department of Physics and Astronomy,
        Aarhus University, DK-8000 Aarhus C, Denmark }

\date{\today}

\begin{abstract}
The lowest-lying states of the Borromean nucleus $^{17}$Ne
($^{15}$O+$p$+$p$) and its mirror nucleus $^{17}$N ($^{15}$N+$n$+$n$)
are compared by using the hyperspheric adiabatic expansion. Three-body
resonances are computed by use of the complex scaling method. The
measured size of $^{15}$O and the low-lying resonances of $^{16}$F
($^{15}$O+$p$) are first used as constraints to determine both central
and spin-dependent two-body interactions. The interaction obtained
reproduces relatively accurately both experimental three-body spectra.
The Thomas-Ehrman shifts, involving excitation energy differences, are
computed and found to be less than 3\% of the total Coulomb energy
shift for all states.  
\end{abstract}


\maketitle

\section{Introduction}

Nuclear halos are expected along the driplines where nucleon single
particle $s$ or $p$-states occur with sufficiently small separation
energy \cite{han95,rii00}. A cluster description of the nucleus
can then be appropriate, being able to visualize the nucleus as a
system made by an ordinary nucleus (the core) surrounded by one or two
nucleons. Since the Coulomb interaction works against
halo formation the appearance of proton halos requires a relatively
light core \cite{rii92,fed94}. Still the degrees of freedom describing
the core and the surrounding protons may decouple and justify a
few-body treatment.

The lightest Borromean proton dripline nucleus is $^{17}$Ne ($^{15}$O$
+ p + p$) which has an odd core mass implying finite core spin. The
low-energy properties of the two-body proton-core subsystems produce
the two sets of known spin-split pairs of resonances \cite{gui98}.
Dealing with the details of such systems is delicate but analogous to
the proper treatment of $^{11}$Li \cite{gar97}.  Unfortunately
the structure of $^{17}$Ne, even within three-body models, seems to
be very controversial and differing rather strongly in available
publications \cite{gri03,tim96}.

Nevertheless $^{17}$Ne was recently discussed \cite{gri02} as an
example revealing new features of the Thomas-Ehrman shift
\cite{ehr51}.  This necessarily introduces the mirror nucleus with
less Coulomb repulsion which then must be more bound and possibly with
a different structure.  In fact the basic assumptions of the
three-body model could be violated. Still it is interesting to push
model applications to test its limits. We shall therefore try to
describe $^{17}$N with precisely the established model parameters and
compare effects of the (lack of) Coulomb interactions.

The paper is organized as follows: In section II we describe very
briefly the method used to compute three-body bound wave functions as
well as three-body resonances. We continue in section III with the
description of the nucleon-nucleon and nucleon-core interactions
needed to study $^{17}$Ne and $^{17}$N.  We then in section IV compare
the properties of the low-lying states of these two mirror nuclei.  In
section V we discuss in details the three-body Thomas-Ehrman shifts
for all these states. We close the paper with a summary and conclusions.

\section{\label{sec2} Three-body method}

The three-body wave functions are computed using the hyperspheric
adiabatic expansion method \cite{fed94,nie01}, that solves the Faddeev
equations in coordinate space. The wave function, $\Psi = \sum
\psi_i$, is then written as a sum of three Faddeev components
$\psi^{(i)}(\bm{x}_i,\bm{y}_i)$ ($i$=1,2,3), where
$\{\bm{x}_i,\bm{y}_i\}$ are the three sets of Jacobi coordinates
defined for instance in \cite{fed94}.  We then introduce the
hyperspheric coordinates, ($\rho=\sqrt{x^2+y^2}$,
$\alpha_i=\arctan({x_i/y_i})$, $\Omega_{x_i}$, and $\Omega_{y_i}$),
and for each value of $\rho$ we expand each component $\psi^{(i)}$ in
terms of a complete set of angular functions:
\begin{equation}
\psi^{(i)}=\frac{1}{\rho^{5/2}} \sum_n f_n(\rho) \phi_n^{(i)}(\rho,\Omega_i);
(\Omega_i\equiv\{\alpha_i, \Omega_{x_i}, \Omega_{y_i} \}), 
\label{eq1b}
\end{equation}

When this expansion is introduced in the Faddeev equations they can
be separated into an angular and a radial part. The angular part takes the form
\begin{equation}
\hat{\Lambda}^2 \phi_n^{(i)}+\frac{2 m \rho^2}{\hbar^2} V_{jk}(x_i)
\left( \phi_n^{(i)} + \phi_n^{(j)}  + \phi_n^{(k)}   \right)  =
\lambda_n(\rho) \phi_n^{(i)}
\label{eq2} 
\end{equation}
where $V_{jk}$ is the two-body interaction between particles $j$ and
$k$, $\hat{\Lambda}^2$ is an angular operator \cite{nie01} and $m$ is
the normalization mass. The complete set of angular functions used in
(\ref{eq1b}) are the eigenvectors of the angular part of
the Faddeev equations, that are labeled with the index $n$ and whose
eigenvalues are denoted by $\lambda_n(\rho)$.

Finally, the coefficients $f_n(\rho)$ in the expansion (\ref{eq1b})
are obtained after solving the coupled set of equations given by the
radial parts of the Faddeev equations:
\begin{eqnarray}
\left[ -\frac{d^2}{d\rho^2} +  \frac{2m}{\hbar^2} (V_{3b}(\rho) - E)
+ \frac{1}{\rho^2}
\left( \lambda_n(\rho)+\frac{15}{4} \right) \right] f_n(\rho) && \nonumber \\
+ \sum_{n'} \left( -2 P_{n n'} \frac{d}{d\rho} - Q_{n n'} \right)f_{n'}(\rho)
= 0 \;\; \;\; && \label{eq3}
\end{eqnarray}
where $V_{3b}$ is a three-body potential used for fine-tuning and the
functions $P_{n n'}$ and $Q_{n n'}$ can be found for instance in
\cite{nie01}.   The eigenvalues $\lambda_n(\rho)$ are essential in
the diagonal part of the effective radial potentials:
\begin{equation}
V_{\mbox{eff}}(\rho)=\frac{\hbar^2}{2m}
    \frac{\lambda_n(\rho)+15/4}{\rho^2}+V_{3b}(\rho)
\label{eq4}
\end{equation}

The hyperspheric adiabatic expansion method was initially designed to
compute three-body bound state wave functions, and therefore the
coupled set of radial equations (\ref{eq3}) was solved for radial
solutions falling off exponentially at large distances. In principle
the method can also be used to calculate continuum and resonance wave
functions. Then we must require the correct asymptotic behaviour for
the solutions to eqs.(\ref{eq3}) as described in \cite{cob97}.

Calculations of resonance wave functions are in practice significantly
simplified by using the complex scaling method \cite{fed03}
where the radial coordinates are rotated into the complex plane by an
arbitrary angle $\theta$.  This transformation of the Jacobi
coordinates ($x\rightarrow x e^{i\theta}$, $y\rightarrow y
e^{i\theta}$) implies that only the hyperradius $\rho$ is transformed
($\rho\rightarrow \rho e^{i\theta}$), while the hyperangles remain
unchanged.  It is then known that as soon as the scaling
angle $\theta$ is larger than the argument of the resonance, then the
complex rotated resonance wave function falls off exponentially,
exactly as a bound state. Therefore, after complex scaling, the
resonances can be computed with the same numerical techniques as for
bound states.

\section{\label{sec3} Two-body potentials}

In this section we investigate the two-body potentials needed to
compute the three-body system $^{17}$Ne ($^{15}$O+$p$+$p$). The
short-range interaction for the mirror nucleus $^{17}$N ($^{15}$N+$n$+$n$)
is then in principle the same although the assumptions of the
three-body model are much less convincing due to the larger binding
energy. The spin-dependence of the effective two-body interactions
must be carefully chosen as shown in \cite{gar03}.  For symmetry
reasons the spin-spin and spin-orbit operators in the nucleon-nucleon
interaction should be $\bm{s}_1\cdot\bm{s}_2$ and
$\bm{\ell}\cdot(\bm{s}_1+\bm{s}_2)$, respectively, where $\bm{s}_1$
and $\bm{s}_2$ are the spins of the two nucleons and $\bm{\ell}$ is their
relative orbital angular momentum.  

For the nucleon-core interaction it is necessary to introduce
operators that conserve the usual mean field quantum numbers, i.e. the
nucleon total angular momentum $\bm{j}_n=\bm{\ell}+\bm{s}_n$ and
the total two-body angular momentum $\bm{j}=\bm{j}_n+\bm{s}_c$, where
$\bm{\ell}$ now is the relative nucleon-core orbital angular momentum,
and $\bm{s}_c$ and $\bm{s}_n$ are the core and nucleon spin,
respectively. This almost uniquely determines the spin operators as the
usual fine and hyperfine terms $\bm{\ell}\cdot\bm{s}_n$ and
$\bm{s}_c\cdot\bm{j}_n$ \cite{gar03}.

\subsection{Nucleon-nucleon interaction}

For the nucleon-nucleon short-range interaction we use the operators
mentioned above, and in particular the potential given in
\cite{gar97}
\begin{eqnarray}
\lefteqn{
V_{NN}(r)=37.05 e^{-(r/1.31)^2}
        -7.38 e^{-(r/1.84)^2} }  \nonumber \\
  &&      -23.77 e^{-(r/1.45)^2}{\bm \ell \cdot \bm s}
        +7.16 e^{-(r/2.43)^2} S_{12}  \label{nnpot} \\
  &&  + \left(49.40 e^{-(r/1.31)^2}+
 29.53 e^{-(r/1.84)^2} \right) {\bm s}_1 \cdot {\bm s}_2   
 + \frac{(e Z_n)^2}{r} \;  \nonumber
\end{eqnarray}
where $\bm{s}=\bm{s}_1+\bm{s}_2$ and $S_{12}$ is the usual tensor
operator, $e$ is the unit electric charge and $Z_n (=0,1)$ is the
nucleon charge number.  The strengths are given in MeV and the ranges
in fm. This potential reproduces the experimental scattering lengths
and effective ranges of the $^1S_0$, $^3P_0$, $^3P_1$, and $^3P_2$
waves. We use the same interaction for relative orbital angular momenta
larger than 1.

\subsection{Nucleon-core interaction}

For the nucleon-core  interaction we construct an
$\ell$-dependent potential of the form:

\begin{eqnarray}
V^{(\ell)}_{N-core}(r) = S_c^{(\ell)} f^{(\ell)}_c(r) &+&S_{ss}^{(\ell)} 
f^{(\ell)}_{ss}(r) 
\bm{s}_c \cdot \bm{j}_n \nonumber \\ -  S_{so}^{(\ell)} 
\frac{1}{r}\frac{d}{dr}
f^{(\ell)}_{so}(r) \bm{\ell} \cdot \bm{s}_{n}
 &+& \frac{Z_c Z_n e^2}{r} {\rm Erf}(r/b_c) \; ,
\label{eq1}
\end{eqnarray}
where $\bm{s}_n$ and $\bm{s}_c$ are the spin of the nucleon and the
core, respectively, $\bm{\ell}$ is the relative orbital angular
momentum between the two particles, $\bm{j}_n=\bm{\ell}+\bm{s}_n$, and
$Z_c$ is the proton number of the core.  The error function ${\rm
Erf}$ describes the nucleon-core Coulomb interaction of a gaussian
core-charge distribution where $b_c = 2.16$ fm is fitted to reproduce
a rms charge radius in $^{15}$O of 2.65 fm. This value is obtained
from the measured rms charge radius in $^{16}$O (2.71 fm \cite{vri87})
by rescaling it by an $A^{1/3}$ factor.

As discussed in \cite{gar03} the choice of these spin operators
permits a clear energy separation of the usual mean-field spin-orbit
partners $\ell_{\ell+1/2}$ and $\ell_{\ell-1/2}$. In this way it is
possible to use a nucleon-core interaction such that the low-lying
states have well defined $\ell_{j_n}$ quantum numbers, like the
$p_{1/2}$ states in $^{10}$Li or $d_{5/2}$ states in $^{16}$F. The use
of the $\bm{s}_c\cdot\bm{s}_n$ spin-spin and
$\bm{\ell}\cdot(\bm{s}_c+\bm{s}_n)$ spin-orbit operators makes this
impossible, since then $j_n$ is not a conserved quantum number and the
states in the two-body system are necessarily mixtures of
$\ell_{\ell+1/2}$ and $\ell_{\ell-1/2}$ components. This is especially
problematic in the case that one of these states is forbidden by the
Pauli principle, like for instance the $p_{3/2}$ waves in $^{10}$Li.
 
The shapes of the central ($f^{(\ell)}_c$), spin-spin
($f^{(\ell)}_{ss}$) and spin-orbit ($f^{(\ell)}_{so}$) radial
potentials in eq.(\ref{eq1}) are chosen to be Woods-Saxon functions,
$1/(1+\exp{((r-b_\ell)/a)})$, with the same diffuseness $a$ in all
cases. Once the range $b_\ell$ of each radial potential is
chosen, the strengths $S_c^{(\ell)}$, $S_{ss}^{(\ell)}$ and
$S_{so}^{(\ell)}$ are adjusted to reproduce the experimental spectrum
of $^{16}$N ($^{15}$N$+n$). For $s$-waves the strengths $S^{(0)}_c$
and $S^{(0)}_{ss}$ are used to fit the energies of the
$s_{1/2}^{(j=0)}$ and the $s_{1/2}^{(j=1)}$ states (0$^-$ and 1$^-$
states). For $d$-waves the strength $S^{(2)}_{so}$ provides an
appropriate spin-orbit splitting of the $d_{3/2}$ and the $d_{5/2}$
states while $S^{(2)}_{c}$ and $S^{(2)}_{ss}$ are used to reproduce
the experimental binding energies of the $d_{5/2}^{(j=2)}$ and the
$d_{5/2}^{(j=3)}$ states (2$^-$ and 3$^-$ states). 

The role of the spin-orbit interaction is here only to place the 
$d_{3/2}$-states relatively high (they must remain unbound), and the 
precise energy of these states is not very relevant.  In any case a 
appropriate estimation of the strength for the spin-orbit interaction 
requires knowledge of the $d_{3/2}^{(j=1)}$ and $d_{3/2}^{(j=2)}$ energies. 
In $^{16}$N there are two unbound 1$^-$/2$^-$ doublets that
could correspond to these states.  Their experimental decay energies
\cite{ajz86} (1.90 MeV and 2.58 MeV, or 2.27 MeV and 2.86 MeV) are
used to estimate the strength of the spin-orbit interaction for
$d$-waves.  

\begin{table}
\caption{\label{tab1} Range ($b_\ell$) and strengths of the central 
($S_s^{(\ell)}$), spin-spin ($S_{ss}^{(\ell)}$), and spin-orbit
($S_{so}^{(\ell)}$) potentials in eq.(\ref{eq1}). The diffuseness $a$
is 0.65 fm in all the cases.}
\begin{ruledtabular}
\begin{tabular}{|c|cccc|}
 $\ell$ & $b_\ell$ (fm) & $S_c^{(\ell)}$ (MeV)  & $S_{ss}^{(\ell)}$ (MeV)  & 
                                      $S_{so}^{(\ell)}$ (MeV$\cdot$fm$^2$)  \\
\hline
 $0$ & 3.00  & $-53.91$ & 0.92 & -- \\
 $1$ & 2.70  & $-19.99$ & 0.69 & $-25.0$ \\
 $1$ & 2.92  & $-54.15$ & 0.35 & $-25.0$ \\
 $2$ & 2.85  & $-58.45$ & 0.24 & $-25.0$ \\
\end{tabular}
\end{ruledtabular}
\end{table}

The value of the range parameter $b_\ell$ is determined by the
fact that by switching on the Coulomb potential the experimental 
spectrum of $^{16}$F ($^{15}$O$+p$) should be reproduced. 
In table \ref{tab1}
we give the resulting values of the parameters used for the
Woods-Saxon radial form factors in eq.(\ref{eq1}).  The partial waves
with $\ell=0$ and $\ell=2$ are by far the most important in the
present context.

The $s$-wave potential has a deeply bound state at $-31.0$ MeV in
$^{16}$N and at $-26.2$ MeV in $^{16}$F.  These states correspond to
the $s_{1/2}$ nucleon states occupied in the $^{15}$N or the $^{15}$O
core. They are then forbidden by the Pauli principle, and should be
excluded from the calculation.  This is implemented as in
refs.\cite{gar97,gar99} by use of the phase equivalent potential which
has exactly the same phase shifts as the initial two-body interaction
for all energies, but the Pauli forbidden bound state is removed from
the two-body spectrum.  We then use the phase equivalent potential of
the central part of the Woods-Saxon $s$-wave potential in table
\ref{tab1}.  Thus the $s$-states actively entering the
three-body calculations are the second states of the Woods-Saxon
potential. For the $d$-states no Pauli exclusion is necessary.

\begin{table}
\caption{\label{tab2} Four lowest states in $^{16}$N and $^{16}$F obtained 
with the nuclear potential specified in table \ref{tab1}. For $^{16}$F
we give the energies and widths of the two-body resonances $(E_R,
\Gamma)$.  The experimental data are from \cite{ajz86}. Error bars are
not specified when they are smaller than the last digit. For unbound
states the energies are decay energies above threshold.}
\begin{ruledtabular}
\begin{tabular}{|c|cc|cc|}
$J^{\pi}$ & $^{16}$N  & Exp.\footnotemark[1]  & $^{16}$F   & Exp.\footnotemark[1]  \\
\hline
 0$^-$ & $-2.37$  & $-2.371$ & (0.53,0.02) & (0.535, $0.040\pm0.020$) \\
 1$^-$ & $-2.09$  & $-2.094$ & (0.71,0.07) & ($0.728\pm0.006$, $<0.040$) \\
 2$^-$ & $-2.49$  & $-2.491$ & (0.96,0.01) & ($0.959\pm0.005$, $0.040\pm0.030$) \\
 3$^-$ & $-2.19$  & $-2.193$ & (1.23,0.01) & ($1.256\pm0.004$, $<0.015$) \\
\end{tabular}
\end{ruledtabular}
\footnotetext[1]{From ref.\cite{ajz86}}
\end{table}

The bound states in $^{16}$N and low-lying resonances in
$^{16}$F with $J^{\pi}$=0$^-$, 1$^-$, 2$^-$, 3$^-$ are all obtained 
by coupling $s_{1/2}$ and $d_{5/2}$ 
with the core-spin of $1/2$. The calculated
results are in table \ref{tab2} compared with the experimental data
for these states.  The procedure of fitting the nuclear potential to
reproduce simultaneously both the $^{16}$N and the $^{16}$F spectra is
apparently efficient as the data is rather nicely reproduced. In fact
this is not possible with other values for $b_\ell$.

In table \ref{tab1} we also specify a $p$-wave interaction although
these partial waves are expected to have only insignificant effects.
The reason is that the lowest $p$-shell is fully occupied in the core
and the unoccupied $p_{3/2}$ orbit is above the $d_{3/2}$-states and
even higher than the $f_{7/2}$.  Nevertheless, since the calculation
will include $p$-wave components at least an estimate of the
parameters for the corresponding interaction is desirable.  We do this
by using the knowledge of the unbound $1^+$ and $2^+$ states in
$^{16}$N immediately above the bound $1^-$ state (with experimental 
decay energies 0.86 MeV and 1.03 MeV \cite{ajz86}, respectively). 
These resonances must arise from the coupling of a $p_{3/2}$ neutron 
with the spin 1/2 of the core (an $f_{7/2}$ neutron can not couple to 1 or 2)
or perhaps by core excitation of a $p_{1/2}$ neutron. Again the 
potential parameters must be such that after switching on the
Coulomb interaction the experimental decay energy of 4.30 MeV for the
$1^+$ state in $^{16}$F has to be also reproduced (the experimental decay
energy of the $2^+$ state of $^{16}$F is not available).
The parameters fulfilling these conditions are given in the second line 
of table \ref{tab1}. The value of $S_{so}^{(1)}$ has been arbitrarily chosen
to be the same as for $d$-waves. 

The lowest $p$ shell is fully occupied in the $^{15}$N or $^{15}$O
core. We should then apply the same treatment as for $s$-waves to the
$p$-wave nucleon-core interaction, using a potential with deeply bound
states that are afterwards removed by the corresponding phase
equivalent potentials. For consistency we also tested a deep $p$-wave 
potential as given in
table \ref{tab1} with range and strengths comparable to the $s$ and $d$ 
potentials.  The bound states in these deep $s$ and $p$ potentials produce
a charge distribution with a rms radius in $^{15}$O of 2.63 fm
consistent with the value used in the Coulomb potential. Furthermore
the binding energies of the $p_{1/2}$ and $p_{3/2}$ states 
in $^{15}$O are $-7.29$ MeV and $-11.3$ MeV, respectively, both consistent 
with the experimental data \cite{ajz91}.

Nevertheless, since the $p$ waves basically have no
effects in the three-body calculation we use for simplification the
shallow $\ell$=1 potential given in table \ref{tab1} without bound
states. In this way the computing time is significantly reduced without
loss in the computations accuracy.

\section{Results for $^{17}$Ne and $^{17}$N}

We use the two-body interactions determined as described in the
previous section. The low-lying nucleon-core valence space is expected
to consist of $s$ and $d$ waves. With spin and parity of $1/2^-$ for
the core and two identical nucleons in the $sd$ valence space we can
construct total angular momentum and parity states with 
$J^{\pi}$=1/2$^-$, 3/2$^-$, 5/2$^-$, 7/2$^-$, and 9/2$^-$. The next shells
($f_{7/2}$, $p_{3/2}$, $\cdots$ ) may also contribute but significant
amounts of such components also indicate similar contributions from
core excitations. These structures involve particle-hole excitations
either from the $sd$ to the $pf$-shell or from the $p$ to the
$sd$-shell. We shall neglect these core excitations.

\subsection{Components}

To solve the eigenvalue problem given in eq.(\ref{eq2}) we expand the
angular eigenvectors in the basis  
$\{ {\cal Y}_{\ell_x \ell_y, L}^{K}(\alpha_i, \Omega_{x_i}, \Omega_{y_i})
\otimes \chi_{s_x s_y, S} \}$, where ${\cal Y}_{\ell_x \ell_y, L}^{K}$ are
the hyperspheric harmonics and $\chi$ is the spin function
\cite{nie01}.  For each of the three Jacobi coordinate sets $i$
the coordinate $\bm{x}_i$ is the vector connecting particles $j$ and
$k$, the quantum number $\ell_x$ is the relative orbital angular
momentum of particles $j$ and $k$, $\ell_y$ is the relative orbital
angular momentum of particle $i$ and the center of mass of the $jk$
two-body system.  The spin $s_x$ is the coupled spin of particles $j$
and $k$, and $s_y$ is the spin of particle $i$.  Finally $L$ and $S$
are the coupling of $\ell_x$ and $\ell_y$, and of $s_x$ and $s_y$,
respectively, and they couple to the total angular momentum $J$ of the
system. The hypermomentum $K$ is given by $2n+\ell_x+\ell_y$ where $n$
is a non-negative integer counting the number of nodes in the Jacobi
polynomials. 

The first step in the calculation is then to choose the components to
be included in the expansion of the angular eigenvectors. By direct
but extensive computations, we have found that the components needed
for $^{17}$Ne are essentially $s$, $p$, and $d$-waves. Only for high
angular momentum ($J$=7/2 and 9/2) higher partial waves can be
relevant. We then use the same components for $^{17}$N.

After solving the angular part of the Faddeev equations (\ref{eq2}) we
extract the angular eigenvalues $\lambda_n(\rho)$, that determine
almost entirely the effective potentials entering in the radial
equations (\ref{eq3}).  For both $^{17}$Ne and $^{17}$N we also here
maintain the same number of lowest-lying adiabatic potentials for
use in the radial equations (\ref{eq3}). We compute first the bound
state solutions falling off exponentially at large distances. Then the
resonance eigenfunctions are found in complete analogy as
exponentially falling solutions to the similar equations obtained by
complex rotation of the hyperradius.

\subsection{Spectrum of $^{17}$Ne}

\begin{figure}
\includegraphics[scale=0.35,angle=-90]{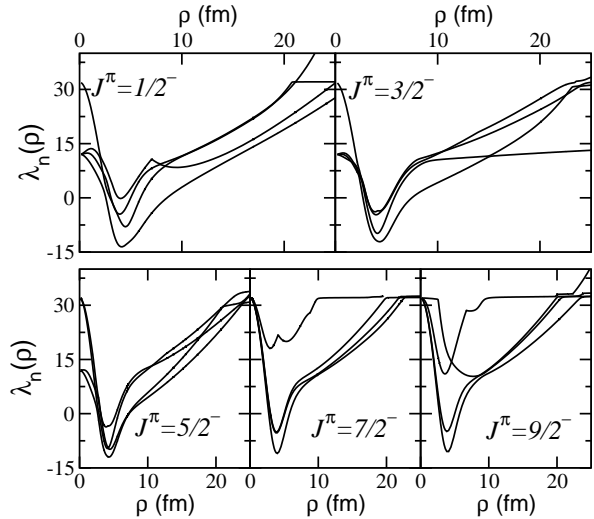}
\caption{\label{fig1} The four lowest angular eigenvalues 
$\lambda_n(\rho)$ for the $1/2^-$, $3/2^-$, $5/2^-$, $7/2^-$ and
$9/2^-$ states of $^{17}$Ne as function of $\rho$ where the
normalization mass $m$ equals the nucleon mass.}
\end{figure}

We show the results in fig.\ref{fig1} for the four deepest effective
potentials for the $1/2^-$, $3/2^-$, $5/2^-$, $7/2^-$, and $9/2^-$
states in $^{17}$Ne.  It is known that at $\rho$=0 the values of the
$\lambda$'s must reproduce the hyperspherical spectrum, $K$($K$+4)
\cite{nie01}.  In our case of positive total parity in the valence
space $K$ must be even, i.e. $K$=0,2,4,$\cdots$.

In the figure we observe that the $\lambda$ function starting at
zero ($K$=0) does not appear.  This is due to the phase equivalent $s$-wave
potential between proton and core where the deepest state of the
initial potential is removed to account for the Pauli principle
\cite{gar99}.  By using the initial deep two-body potential instead we 
obtain a $\lambda$ function starting at zero for $\rho=0$ and
diverging to $-\infty$ at large distances. This behaviour of the
lowest $\lambda$ characterizes the existence of a bound two-body state
\cite{nie01}.  This state is actually the Pauli forbidden state which 
could have been computed and then omitted from the basis. Instead we
suppressed the Pauli forbidden state by using the more consistent
procedure with the phase equivalent potential.

For short-range potentials it is also known that at infinity the
values of the $\lambda$'s must again follow the hyperspherical
spectrum \cite{nie01}. However, this behaviour is changed for
eigenvalues corresponding to unbound two-body states as soon as
long-range interactions like the Coulomb potential are present.  The
reason is that the influence of the short-range interactions then
disappear outside $\rho$-values corresponding to a few times the range
of the interaction whereas the Coulomb potentials multiplied by
$\rho^2$ give rise to linearly increasing $\lambda$ functions even at
asymptotically large distances \cite{fed96}. This linear increase must
appear as soon as only the Coulomb potential has an influence. The
slopes depend on the geometric structure of the three-body system as
the size increases.

The ground state of $^{17}$Ne is bound, and has a two-proton
separation energy of $-944$ keV. The structure is about equal amounts
of proton-core $s^2$ and $d^2$-waves.  The computed root mean square
radius is 2.8 fm consistent with the experimental value of
$2.75\pm0.07$ fm \cite{oza94}.  All the excited states are unbound and
computed by application of the complex scaling method.  The excitation
energies of the two lowest excited states are 1288$\pm$8 keV for the
$3/2^-$ state and 1764$\pm$12 keV for the $5/2^-$ state
\cite{gui98}. For both these states proton-core $s$-$d$ mixed components
are dominating.  In \cite{gui98} also a $7/2^-$ and a $9/2^-$ states
are reported with excitation energies 2997$\pm$11 keV and 3548$\pm$20 keV,
respectively.  These four excitation energies correspond to the decay
energies (energies above threshold) given in the second column of
table~\ref{tab3}.  For these states the $d^2$-waves dominate.
More details about the structure is available in \cite{gar03a}.

The resonances obtained for $^{17}$Ne are extremely narrow with widths
much smaller than the accuracy of our calculations.  Thus, application
of the complex scaling method allows the use of very small scaling
angles.  Typically complex scaling angles of $\theta$=10$^{-5}$ are
able to find the $^{17}$Ne resonances.  For these scaling angles the
complex scaled $\lambda$'s can hardly be distinguished from the
non-rotated functions in fig.\ref{fig1}.  The imaginary parts are very
small and would appear on the zero line if plotted on the figure.

\begin{table}
\caption{\label{tab3} The second and third columns give the experimental 
and computed bound state ($1/2^-$) and decay energies ($3/2^-$, 
$5/2^-$, $7/2^-$, and $9/2^-$) in $^{17}$Ne (in MeV). The fourth column are
the strengths $S$ (in MeV) of the gaussian three-body forces
that for a range of 4.0 fm give rise to  
energies matching the experimental values. The fifth column gives 
the expectation value of the three-body force for the corresponding 
$^{17}$Ne solutions. The last column is the contribution to the norm
of the first three terms in the expansion (\ref{eq1b}).}
\begin{ruledtabular}
\begin{tabular}{|c|cc|c|c|c|}
$J^{\pi}$ & $E_{exp}$ &$E_{comp}$ &$S$  
                            & $\langle V_{3b} \rangle$ &
      $\lambda_{n=1,2,3}$ (\%) \\
\hline
$1/2^-$ & $-0.94$ & $-0.79$ & $-0.6$  & $-0.2$ &  88.5, 11.1, 0.4 \\
$3/2^-$ & $ 0.34$ & $ 0.63$ & $-1.4$  & $-0.3$ &  90.7, 8.9, 0.2 \\
$5/2^-$ & $ 0.82$ & $ 0.91$ & $-0.4$  & $-0.1$ &  77.2, 16.9, 5.6 \\
$7/2^-$ & $ 2.05$ & $ 2.24$ & $-0.8$  & $-0.2$ &    97.5, 2.3, 0.2 \\
$9/2^-$ & $ 2.60$ & $ 2.70$ & $-0.1$  & $-0.1$ &    91.9, 4.2, 3.8 \\
\end{tabular}
\end{ruledtabular}
\end{table}

Using the $\lambda$ functions in fig.\ref{fig1} we obtain the
$^{17}$Ne ground state binding energy and the decay energy of the
excited states shown in the third column of table~\ref{tab3}. As seen
in the table the computed states are systematically less bound than
the experimental value. This fact is actually expected, since
three-body calculations using pure two-body interactions typically
underbind the system. This problem is solved by inclusion of the weak
effective three-body potential $V_{3b}$ in (\ref{eq3}), that accounts
for three-body polarization effects arising when the three particles
all are close to each other.  Therefore the three-body potential has
to be of short range, while the three-body structure essentially is
independent of the precise shape.  This construction furthermore
ensures that the two-body resonances remain unaffected within the
three-body system after this necessary fine-tuning.  The effective
total potential entering is then given by (\ref{eq4}).

The precise range of the three-body interaction also plays a limited
role.  This is because the three-body force is very weak compared to
the depth of the full potential and furthermore it is largest for
small $\rho$-values, where the total potential is highly repulsive.
It is then clear that the main structure of the system can not be
significantly modified by the choice of one or another of such
three-body interactions.  In table \ref{tab3} we give the strengths of
the gaussian three-body potentials which for range equal to 4 fm are
needed to match the experimental energies of all these (ground and)
excited states.  One way to measure and compare the effect of the
three-body force in the different calculations is to compute the
expectation value $\langle V_{3b}(\rho) \rangle$ of the three-body
potential for the corresponding $J^\pi$ solutions.  In table
\ref{tab3} we also give this quantity which measures the contribution
of the three-body force to the energy of the three-body system.  A
variation of the range of the three-body force within reasonable
limits is not modifying the results.

In the last column of table~\ref{tab3} we give the contribution to the 
norm of the wave function of the first three terms in the expansion 
(\ref{eq1b}). Typically only two terms are enough to get an accuracy
of 99\%, and only for the $5/2^-$ and $9/2^-$ states the third term
is giving a sizable contribution.

The spectrum of $^{17}$Ne has been previously investigated 
in \cite{gri03,tim96}. In both works the 3/2$^-$ 
and 5/2$^-$ levels are reversed compared to the experimental data,
although in \cite{gri03} this deficiency is corrected by use of
an appropriate three-body interaction. In the present work these 
problems are not encountered.  When only
the two-body forces describing properly the $^{16}$F spectrum are used,
the ordering in the computed $^{17}$Ne spectrum is correct, as seen in 
the third column of table~\ref{tab3}. Then,  the use of a small effective 
three-body force is enough to fit the experimental data.

\subsection{Spectrum of $^{17}$N}

Interchanging all neutrons and protons in $^{17}$Ne leads to the
mirror system $^{17}$N which then analogously should be described as a
three-body system with the $^{15}$N-core surrounded by two neutrons.
The structure should then be obtained simply by switching off the
Coulomb interaction for $^{17}$Ne, since the strong interaction is
precisely the same due to charge symmetry. In this way we can compute
the properties of $^{17}$N. 

The results are listed in column two of table \ref{tab4} and
not surprisingly stronger binding is obtained. First, $^{17}$N is not
a Borromean system.  The number of three-body bound states has also
increased substantially, i.e. we find two bound states both for
$1/2^-$ and $3/2^-$, and one for $5/2^-$, $7/2^-$ and $9/2^-$.  The
computed energies of these states agree pretty well with the
experimental values \cite{ajz86}.  The discrepancy with the experiment
is always smaller than 5\%. Only for the excited $1/2^-$ and $3/2^-$
states a larger disagreement with the measured energies appears.  In
the third column of the table the calculations labeled by A+WS use 
the accurate Argonne nucleon-nucleon potential denoted in \cite{wir95} 
by $v_{18}$ and the Woods-Saxon nucleon-core potential in table~\ref{tab1}.

\begin{table}
\caption{\label{tab4} $^{17}$N spectrum obtained using the same nuclear
two-body interactions as for $^{17}$Ne, and the effective three-body
force given in table \ref{tab3}. The third column gives the results
obtained with the Argonne (A) nucleon-nucleon potential plus the 
Woods-Saxon (WS) nucleon-core interaction (table~\ref{tab1}).  
The last column gives the experimental data \cite{ajz86}. }
\begin{ruledtabular}
\begin{tabular}{|c|ccc|}
$J^{\pi}$ &$E$ (MeV)  &$E_{\mbox{\scriptsize A+WS}}$ (MeV)& Exper. \\
\hline
 $1/2^-$ & $-8.54$  & $-8.31$& $-8.374$ \\
         & $-3.72$  & $-3.66$& $-4.711$ \\
 $3/2^-$ & $-6.63$  & $-6.80$& $-7.000$ \\
         & $-3.83$  & $-5.03$& $-5.174$ \\
 $5/2^-$ & $-6.32$  & $-6.36$& $-6.467$ \\
 $7/2^-$ & $-5.24$  & $-5.17$& $-5.245$ \\
 $9/2^-$ & $-4.58$  & $-4.59$& $-4.745$ \\
\end{tabular}
\end{ruledtabular}
\end{table}

\begin{figure}
\includegraphics[scale=0.35,angle=-90]{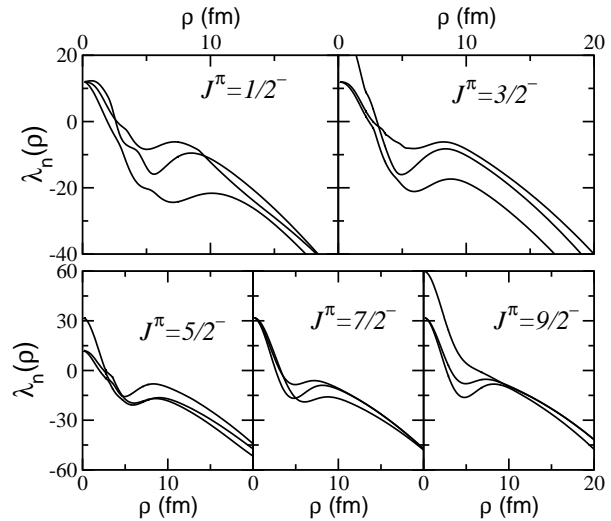}
\caption{\label{fig2} The three lowest effective potentials $\lambda_n(\rho)$
for the $1/2^-$ and $3/2^-$ (upper part), and $5/2^-$, $7/2^-$, and $9/2^-$ 
(lower part) states of $^{17}$N.}
\end{figure}

The relatively small differences between the mirror nuclei are perhaps
not as self-evident if we consider the behavior of the angular
eigenvalues shown in fig.\ref{fig2}. We plot only the three lowest
functions used to compute the different states in $^{17}$N.  As for
$^{17}$Ne there is no $\lambda$ function starting at zero due to the
removal of the Pauli forbidden $s$-state by use of the phase
equivalent potential.  The main difference compared to fig.\ref{fig1}
is the divergence towards $-\infty$ of all the $\lambda$ functions.
This is a reflection of corresponding bound states in the two-body
subsystems consistent with the quantum numbers of the three-body
system. This underlines that $^{17}$N cannot be a Borromean nucleus
\cite{nie01}.  The qualitatively different behavior seen in
figs.\ref{fig1} and \ref{fig2} also emphasizes that the agreement with
measurements for both nuclei is not a trivially build-in property of
the present description. The model is consistent in a more profound
way.

\begin{figure}
\includegraphics[scale=0.35,angle=-90]{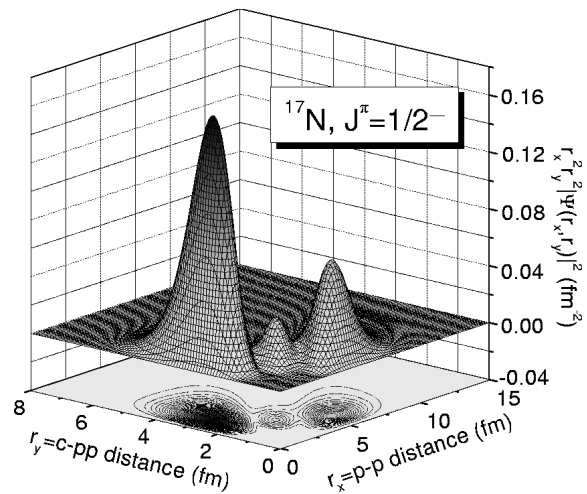}
\caption{\label{fig3} Contour diagram for the probability distribution 
of the $1/2^-$ ground state of $^{17}$N. The square of the three-body
wave function is integrated over the directions of the two Jacobi 
coordinates. }
\end{figure}

The spatial structure of $^{17}$N is less extended than that of
$^{17}$Ne because its larger binding energy. The length scale as
defined in \cite{jen03} is $\rho_0\approx 5.5$ fm ($\rho_0\approx
5$ fm for $^{17}$Ne) and the dimensionless measures of size and
binding energy are $\langle(\rho / \rho_0)^2\rangle \approx 0.9$ and
$m B \rho_0^2 /\hbar^2 \approx 6.0$. The ground state of $^{17}$N is
located in the same region as ordinary nuclei. Among the excited
states shown in table \ref{tab4} the less bound is the second $1/2^-$
state. For this state the corresponding dimensionless size and binding
are 1.9 and 3.4, respectively, still located in the region of ordinary
nuclei. The three-body structure is further illustrated in
fig.\ref{fig3}, where we show the square of the three-body wave
function integrated over the directions of the Jacobi coordinates and
multiplied by the volume element. The structure resembles that of
$^{17}$Ne with two similar dominating peaks.

The known properties of $^{17}$N seem to be rather well reproduced
with the model parameters for $^{17}$Ne. An additional very small
retuning of the strength of the effective three-body interaction could
fit the lowest of each $J^\pi$ states. However, this would not improve
the agreement of the second $1/2^-$ and $3/2^-$ states which then move
into the positions $-3.62$ MeV and $-4.14$ MeV, respectively.  This is
probably because other effects are important, e.g. different
components could now contribute both from valence space and from
excitations. This is equivalent to an attempt to describe higher-lying
resonances in $^{17}$Ne. They may also require another and perhaps
enlarged Hilbert space.  Further investigations of these well bound
excited states are beyond the scope of the present report.

\section{Thomas-Ehrman shifts}

The only difference in the computations of the two mirror nuclei is
omission of the Coulomb interaction for $^{17}$N. This similarity is
assumed inherently to describe the fundamental charge symmetry of the
strong interaction.  The immediate implication is that the differences
in the spectra entirely must be produced by the Coulomb potential. The
obvious difference is the shift of all energies towards stronger
binding when the Coulomb potential is suppressed. However, this trivial overall
shift is accompanied by a modified structure of the states. This is
especially seen for the $s$-wave components which are less influenced
by centrifugal barrier effects. The Coulomb repulsion tends to
increase the size of a given state simply by minimizing the energy.  The
$s$-waves are here more influenced than higher partial waves and the
shifts are then larger. This double difference in energy
(excitation energy difference) is for single-particle energies called
the Thomas-Ehrman shift \cite{ehr51}.  It is due to the
Coulomb interaction but not necessarily in a straightforward way.  In
a recent work \cite{gri02} the Thomas-Ehrman shift was investigated in
the three-body systems $^{12}$O and $^{16}$Ne.

\begin{table}
\caption{\label{tab5} The first column indicates the different states
and calculations performed. The initials refer to the 
nucleon-nucleon+nucleon-core interactions used (G: Gaussian, WS: Woods-Saxon,
A: Argonne). The second column gives the 
measured excitation energies of low-lying states in $^{17}$Ne.
The computed $^{17}$Ne spectrum is identical. The measured and
computed ($^{17}$N)$_{th}$ spectra in the mirror nucleus $^{17}$N are
given in the third and fourth columns. The fifth and sixth columns
contain the experimental ($\Delta_e$) and computed ($\Delta_{th}$)
Thomas-Ehrman shifts. The seventh column shows the difference
$\Delta_{e-th}$ between experimental and computed energies for
$^{17}$N (see table \ref{tab4}). All the energies are given in MeV.}
\begin{ruledtabular}
\begin{tabular}{|l|cccccc|}
  $\hspace*{8mm}J^\pi$ & $^{17}$Ne  &  $^{17}$N & ($^{17}$N)$_{th}$ & 
                       $\Delta_e$ & $\Delta_{th}$  &  $\Delta_{e-th}$\\
\hline
$1/2^-$ G+WS  & 0.0 & 0.0 & 0.0 & 0.0 & 0.0 & 0.17 \\
$1/2^-$ A+WS   & " & " & " & " & " & $-0.06$ \\
\hline
$3/2^-$ G+WS  & 1.29 & 1.37 & 1.91 & 0.08 & 0.62 & $-0.37$\\
$3/2^-$ A+WS   & " & " & 1.51 & " & 0.22 & $-0.20$ \\
\hline
$5/2^-$ G+WS  & 1.76 & 1.91 & 2.22 & 0.15 & 0.46 & $-0.15$ \\
$5/2^-$ A+WS   & " & " & 1.95 & " & 0.19 & $-0.11$ \\
\hline
$7/2^-$ G+WS & 3.00 & 3.13  & 3.30 & 0.13 & 0.30 & $-0.01$ \\
$7/2^-$ A+WS & " & "  & 3.14 & " & 0.14 & $-0.08$ \\
\hline
$9/2^-$ G+WS & 3.55 & 3.63  & 3.96 & 0.08 & 0.41 & $-0.17$ \\
$9/2^-$ A+WS & " & "  & 3.72 & " & 0.17 & $-0.16$ \\
\end{tabular}
\end{ruledtabular}
\end{table}

We compare in table~\ref{tab5} the spectrum of excitation energies
for the two mirror nuclei. The second and third columns give the
experimental excitation energies for the different states in $^{17}$Ne
and $^{17}$N, respectively. The gaussian nucleon-nucleon interaction 
in eq.(\ref{nnpot}) and the Woods-Saxon nucleon-core potential in 
table~\ref{tab1} together with the three-body forces in table \ref{tab3}
reproduce the experimental $^{17}$Ne excitation energies. This calculation
is denoted by G+WS. As in table \ref{tab4} 
the calculation denoted by A+WS uses the Argonne nucleon-nucleon potential
and the Woods-Saxon nucleon-core potential in table~\ref{tab1}.
A small three-body force also permits to reproduce the experimental
$^{17}$Ne spectrum.

When these interactions are used for $^{17}$N the energies in table
\ref{tab4} and in the fourth column of table \ref{tab5} are
obtained. The computed excitation energies $(^{17}\mbox{N})_{th}$ of
$^{17}$N are systematically higher than those measured. Still the
agreement is surprisingly good in view of the fact that each state is
computed independently by expansion on individual basis components
without any parameter adjustment. This agreement is especially good
for the A+WS calculation, where the short distance properties of the
nucleon-nucleon interaction are carefully treated. Furthermore the
states in $^{17}$N are well bound and the assumptions of independent
degrees of freedom in the three-body cluster model cannot be very well
fulfilled.

The experimental shifts ($\Delta_e$) are given in the fifth column of
table \ref{tab5}. The total Coulomb shift (given by the energy difference
between a $^{17}$Ne state and the corresponding $^{17}$N state) is
of around 7.4 MeV. The values of $\Delta_e$ are then remarkably small 
compared to the 10\% of the total Coulomb shift for the classical example
of the single-particle $s$ and $d$-states in $^{17}$O and $^{17}$F. It
is tempting to conjecture that this is due to the stronger effect of
the centrifugal barrier in the three-body system where even the
$s$-states feel a barrier. The Coulomb repulsion effect 
is then less pronounced than for a two-body system where the absence
of the centrifugal barrier is the basic explanation.

The computed shifts ($\Delta_{th}$) are shown in the sixth column of
the table. For the G+WS calculation they are clearly larger than the 
measured values, but also in these cases the computation represents an 
accuracy better than 10\% of the Coulomb shift. In the A+WS
computation $\Delta_{th}$ is no more than 3\% of the Coulomb shift,
and shows a better agreement with the experiment.

However, the $(^{17}\mbox{N})_{th}$ values given in the fourth column
of table \ref{tab5} are obtained by comparison to the computed ground
state $1/2^-$ energy for each calculation. This energy differs
from the experimental value by 170 keV and $-60$ keV for the
G+WS and A+WS calculations, respectively (see last column in
table~\ref{tab5}).  These numbers do not enter when comparing to the
total two-nucleon separation energy, but they do enter in the computed
shifts ($\Delta_{th}$) in the sixth column of table~\ref{tab5}.
Therefore the uncertainty reflected in these computed ground state
energies is comparable to the experimental Thomas-Ehrman shift
$\Delta_e$ we are trying to reproduce.  A dedicated effort is needed
to reduce these uncertainties.

In these comparisons the measured values include all many-body effects
while the computations are within the three-body model. To
estimate effects of structure changes we can compare
 the properties of these mirror nuclei through direct computations
of Coulomb energies with the model wave functions.  Following
\cite{aue00} we consider the first-order perturbative contribution to
the $^{17}$N energy from the Coulomb potential, i.e.,
\begin{equation}
\Delta_c^{(1)}=\langle \Psi(^{17}\mbox{N})|V_{coul}|\Psi(^{17}\mbox{N}) \rangle
\label{eq6}
\end{equation}
where $\Psi$ is the three-body $^{17}$N wave function obtained without
Coulomb interaction between core and valence particles and
reproducing the experimental $^{17}$N spectrum. Then the
valence neutrons are substituted by protons in precisely the same
configurations arriving at an artificial $^{17}$Ne wave function. Then
$V_{coul}$ is the resulting Coulomb interaction between the three
pairs of charged particles.  Thus $\Delta_c^{(1)}$ is the diagonal
contribution to the Coulomb shift if the wave function remains
unchanged.  In \cite{aue00} the difference 
\begin{equation}
\Delta_{TE}=\Delta_c-\Delta_c^{(1)}
\end{equation}
is referred to as the Thomas-Ehrman shift, where $\Delta_c$ is the
experimental shift between the $^{17}$N and $^{17}$Ne energies.  Then
$\Delta_{TE}$ represents the reduction in the Coulomb energy in
$^{17}$Ne produced by the modified structure in the single particle
states. 

In table~\ref{tab6} we give these Thomas-Ehrman shifts ($\Delta_{TE}$)  
arising from
experimental ($\Delta_c$) and computed ($\Delta_c^{(1)}$) Coulomb shifts 
between the mirror nuclei $^{17}$Ne and $^{17}$N. We also give
$\Delta_{S}$, that is an estimate of the Coulomb shifts 
due to changes of structure included in the three-body model.
Again we give the results for the G+WS and A+WS. All the $\Delta_{TE}$ 
are less than 3$\%$ of the diagonal Coulomb shift. 

\begin{widetext}
\begin{longtable}{lcccccccccc}
\caption{\label{tab6} Experimental Coulomb shift ($\Delta_c$),
first order contribution ($\Delta_c^{(1)}$, eq.(\ref{eq6})),
Thomas-Ehrman shift ($\Delta_{TE}$) and $\Delta_{S} \equiv \Delta_{TE}
- \Delta_{e-th}$ for the $1/2^-$, $3/2^-$, $5/2^-$, $7/2^-$, and
$9/2^-$ states. G+WS and A+WS refer to
the nucleon-nucleon+nucleon-core interactions used in the different
calculations (G: Gaussian, WS: Woods-Saxon, A: Argonne). All the energies 
are given in MeV. } \\ \hline\hline
\begin{tabular}{|l|cc|cc|cc|cc|cc|}
  $J^\pi \rightarrow$
 & \multicolumn{2}{c|}{$1/2^-$} & \multicolumn{2}{c|}{$3/2^-$}
 & \multicolumn{2}{c|}{$5/2^-$} & \multicolumn{2}{c|}{$7/2^-$}  &
   \multicolumn{2}{c|}{$9/2^-$} \\
\hline
 &  G+WS & A+WS & G+WS & A+WS & G+WS & A+WS &
    G+WS & A+WS & G+WS & A+WS   \\ \hline
$\Delta_c$ &  \multicolumn{2}{c|}{7.43} & \multicolumn{2}{c|}{7.34}
 & \multicolumn{2}{c|}{7.29} & \multicolumn{2}{c|}{7.30}  &
   \multicolumn{2}{c|}{7.35}  \\
$\Delta_c^{(1)}$ & 7.64 & 7.53 & 7.45 & 7.60 & 
                   7.44 & 7.51 & 7.58 & 7.56 & 
                   7.53 & 7.53  \\
$\Delta_{TE}$ & $-0.21$ & $-0.10$ & $-0.11$ & $-0.26$  
              & $-0.15$ & $-0.22$ & $-0.28$ & $-0.26$ 
              & $-0.18$ & $-0.18$  \\
$\Delta_{S}$  & $-0.38$ & $-0.04$ & $+0.26$ & $-0.06$ 
              & $ 0.00$ & $-0.11$ & $-0.27$ & $-0.18$ 
              & $-0.01$ & $-0.02$  \\ \hline\hline
\end{tabular}
\end{longtable}
\end{widetext}

The values of $\Delta_{TE}$ obtained are again highly influenced by the 
structure
of the $^{17}$N states with the different calculations. As mentioned
above, the agreement between computed and experimental two-neutron
separation energies in $^{17}$N can be considered rather good 
(see table \ref{tab4}).
The experimental energy is recovered for the calculations in
table~\ref{tab6} by including in each case the appropriate three-body
interaction, that as we know, keeps almost unchanged the three-body
structure. From table~\ref{tab4} we observe that in some cases the
$^{17}$N states are up to 0.2 MeV more bound in one of
the computations compared to the other. These states are then
more compact, and the Coulomb repulsion should in principle be larger.
This is clearly seen in table~\ref{tab6}, where the larger values
for $\Delta_c^{(1)}$ for the 1/2$^-$, 3/2$^-$, 5/2$^-$, 7/2$^-$, and 
9/2$^-$ states are respectively the G+WS, A+WS, A+WS, G+WS, and A+WS
calculations,
that are precisely the computations giving the more bound state
for each level. In the 9/2$^-$ case, since the binding energy 
with the G+WS and A+WS calculations es pretty much the same, then  
also the $\Delta_c^{(1)}$ value is the same in both cases.

The main conclusion after analysis of tables \ref{tab5} and
\ref{tab6} is that the computed Thomas-Ehrman shifts are highly
determined by the detailed structure of the $^{17}$N states (for
instance different three-body forces can significantly change the
results).  The small change in
the structure from calculation to calculation is in our case important
enough to produce large uncertainties in the computed Thomas-Ehrman
shifts.  These uncertainties are probably much smaller for a system,
which in contrast to $^{17}$N, undoubtedly can be described as a
three-body system.

\section{Summary and conclusions}

The Borromean nucleus $^{17}$Ne and its non-Borromean mirror $^{17}$N
are investigated in a three-body model where two nucleons surround
cores of $^{15}$O and $^{15}$N, respectively. We employ the well
tested hyperspheric adiabatic expansion method. Then the two-body
interactions must first be determined to reproduce the properties of
the two-body subsystems. We carefully choose a spin-dependent form of
the nucleon-core interaction such that the orbits of both the core and
the valence nucleons can be treated consistently to lowest order in
the mean-field approximation. Then we are guarantied that the
fundamental assumption in the three-body model of decoupled motion of
core and valence nucleons is fulfilled as well as possible. We then
proceed to determine parameters of the interactions such that the
lowest four resonance energies of $^{16}$F are reproduced. For this we
use the Coulomb energy of a gaussian charge distribution of measured
root mean square radius. The computed rms radius of the core is in
agreement with the measured size.

The three-body ground state and four measured excited states of
$^{17}$Ne are then computed. The computed states are systematically
slightly underbound compared to the experimental energies, but
reproducing properly the experimental angular momentum ordering. 
Agreement with the experimental energies is obtained by use of 
a weak attractive short-range effective three-body interaction. 

We then turned to the mirror nucleus $^{17}$N which is well bound and
with a number of bound excited states. They are also computed in the
three-body model although the basic assumptions can not be expected to
hold. Still the energies are close to the observed values. We
therefore continued to compute the Coulomb energy and the three-body
Thomas-Ehrman shifts, which are as double energy differences very
sensitive to inaccuracies and model assumptions.  In general
sufficient accuracy can not be reached within three-body models
applied to well bound systems like $^{17}$N. The reason is that
neglected degrees of freedom now can contribute with similar small
amounts.

In conclusion, the three-body model describes efficiently the cluster
structure of $^{17}$Ne and in addition also surprisingly well the well
bound mirror nucleus $^{17}$N. The computed three-body Thomas-Ehrman
shifts are then meaningful although relatively inaccurate.

\begin{acknowledgments}
We want to thank Hans Fynbo for useful suggestions and discussions.
\end{acknowledgments}

\end{document}